\documentclass[sigconf]{acmart}

\AtBeginDocument{%
  }

\acmConference[Conference acronym 'XX]{Make sure to enter the correct
  conference title from your rights confirmation email}{June 03--05,
  2018}{Woodstock, NY}
  
\usepackage{cleveref}

\begin{document}

\title{VidAnimator: User-Guided Stylized 3D Character Animation from Human Videos}

\author{Xinwu Ye}
    \authornote{Both authors contributed equally to this research.}
    \email{xwye23@m.fudan.edu.cn}
    \orcid{0009-0008-3164-0373}
\author{Jun-Hsiang Yao}
    \authornotemark[1]
    \email{rxyao24@m.fudan.edu.cn}
    \orcid{0009-0000-8944-1942}
    
\affiliation{%
  \institution{School of Data Science, Fudan University}
  \city{Shanghai}
  \country{China}
}

\author{Jielin Feng}
    \affiliation{%
      \institution{Fudan University}
      \streetaddress{}
      \city{}
      \country{China}}
\email{jielinfeng23@m.fudan.edu.cn}
\orcid{0009-0008-1943-5609}

\author{Shuhong Mei}
    \affiliation{%
      \institution{School of Data Science, Fudan University}
      \city{Shanghai}
      \country{China}
    }
\email{1010647987@163.sufe.edu.cn}
\orcid{0009-0005-2282-5403}

\author{Xingyu Lan}
    \affiliation{%
      \institution{School of Data Science, Fudan University}
      \city{Shanghai}
      \country{China}
    }
\email{xingyulan96@gmail.com}
\orcid{0000-0001-7331-2433}

\author{Siming Chen}
    \affiliation{%
      \institution{School of Data Science, Fudan University}
      \city{Shanghai}
      \country{China}
    }
\email{simingchen@fudan.edu.cn}
\orcid{0000-0002-2690-3588}

\renewcommand{\shortauthors}{Ye and Yao, et al.}

\begin{abstract}
    With captivating visual effects, stylized 3D character animation has gained widespread use in cinematic production, advertising, social media, and the potential development of virtual reality (VR) non-player characters (NPCs). 
    However, animating stylized 3D characters often requires significant time and effort from animators. 
    We propose a mixed-initiative framework and interactive system to have stylized 3D characters mimic motion in human videos. 
    The framework takes a single-view human video and a stylized 3D character (the target character) as input, captures the motion of the video, and then transfers the motion to the target character. In addition, it involves two interaction modules for customizing the result.
    Accordingly, the system incorporates two authoring tools that empower users with intuitive modification.
    A questionnaire study offers tangible evidence of the framework's capability of generating natural stylized 3D character animation similar to motion of the video.
    Additionally, three case studies demonstrate the utility of our approach in creating diverse results. 
\end{abstract}


\begin{teaserfigure}
  \includegraphics[width=\textwidth, height=0.549\textwidth]{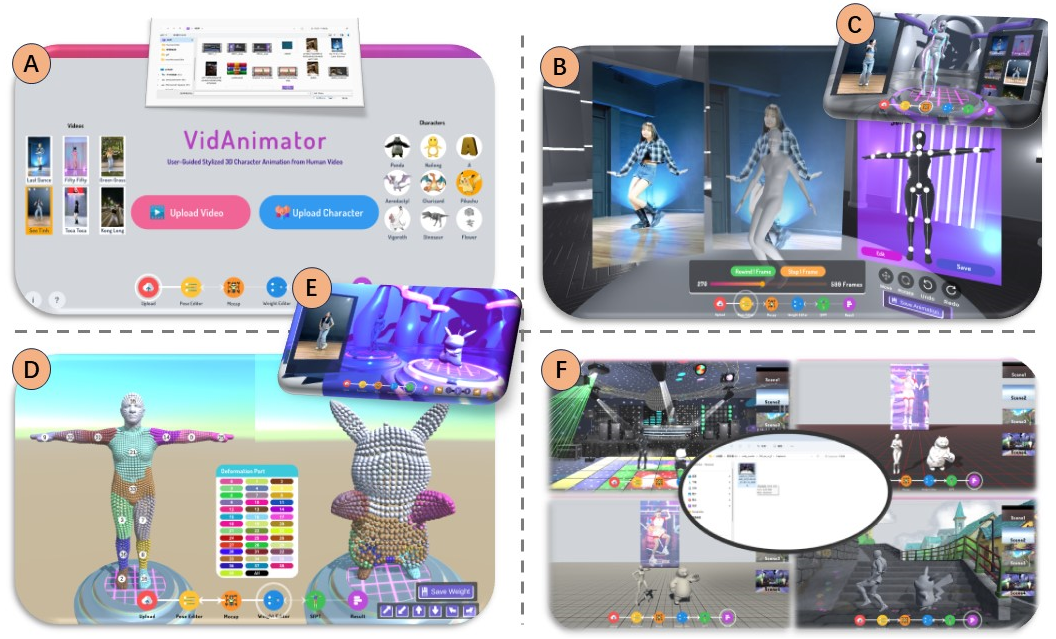}
  \caption{VidAnimator consists of (A) the Upload Interface for video and 3D character; (B) the Pose Editor and (C) MoCap Interface, facilitating keyframe selection and joint position/rotation editing with real-time previews from Video-to-Human; (D) the Skinning-Weight Editor and (E) MoTrans Interface, enabling weight adjustments and previews from Character-to-Stylized 3D Model; and (F) the Results Interface, allowing scene selection and the integration of animated characters, with video capture capabilities for video output.}
  \Description{}
  \label{fig:teaser}
\end{teaserfigure}

\maketitle
\section{Introduction}\label{sec-1}

In recent years, the rapid development of stylized 3D character animation has revolutionized the way we interact with virtual characters. Stylized 3D characters encompass both humanoid and non-humanoid forms. Humanoid 3D characters have body proportions and topologies akin to humans, though they may include additional features like wings or tails. In contrast, non-humanoid 3D characters, such as quadrupedal animals or even stylized flowers, possess body proportions and topologies distinct from humans. Humanoid 3D characters enable anthropomorphism, which means giving them human-like attributes and movements. This enrichment enhances their lifelike qualities, making them capable of mimicking human actions, expressions, and emotions. This advancement in technology finds applications across diverse creative industries, including filmmaking, video games, advertising, and virtual/augmented reality. Aside from specialized fields, growing social networks and faster information sharing are driving greater public demand for this technology \cite{zhaoMetaversePerspectivesGraphics2022}. An increasing number of general users are expressing interest in customizing their own 3D character animations for sharing on social media platforms like TikTok. The applications of this technology are continually expanding, making it an indispensable tool in various fields. However, beneath the visually captivating surface of these complex creations, there exist a series of challenges and limitations that warrant exploration.

A key challenge of stylized 3D character animation is its inherent intricacy, which complicates the animation process.
Although mature tools \cite{Maya2019, Blender3DModelling2018} are available for animating stylized 3D characters, they invariably require a high level of expertise in various facets, such as rigging, weight painting, keyframe animation, as well as the finer aspects of timing and pacing, placing a substantial burden on animators. 
Mastering these technical skills is challenging, and employing them to create stylized 3D character animations is both time-consuming and labor-intensive, especially given the extensive design space. 
Furthermore, producing high-quality stylized 3D animations requires a deep understanding of animation principles, including aspects like character movement, poses, and center of gravity, which are not mastered by general users. 
Driven by these challenges, we are motivated to develop an automated approach for creating stylized 3D character animations.

Presently, a myriad of motion capture tools leveraging human videos exists to automate the creation of 3D character animations \cite{choiStaticFeaturesTemporally2021, wandtCanonposeSelfsupervisedMonocular2021, liLearningHumanDynamics2022}. However, these methods primarily focus on animating 3D human characters, leaving stylized 3D characters—an equally important category—somewhat neglected. To fill this gap, we aim to automate the process of animating stylized 3D characters based on human videos.
We have conducted multiple experiments using the latest DeepMotion MoCap Model for motion capture and Skeleton-Free Pose Transfer (SFPT) for motion transfer. Through extensive experiments, we have identified certain concerns. (1) There are inaccuracies when using DeepMotion MoCap Model. For instance, when multiple individuals or body parts occlude or overlap with each other, the results of MoCap model may not match the motion in videos perfectly. (2) The implicit correspondence between the 3D human character and the target character predicted by the SFPT does not always meet users' expectations.

To address these deficiencies, we propose an innovative framework that incorporates motion capture (MoCap), motion transfer (MoTrans), and interaction module consisting of a pose editor and a weight editor to automate  stylized 3D character animation from human videos interactively. In the MoCap phase, we employ the DeepMotion MoCap model \cite{DeepMotion2023} to capture motion from user-input human videos, which is then applied to a 3D human character. Subsequently, the captured motion of the 3D human character is transferred to the user-input stylized 3D character (the target character) based on Skeleton-Free Pose Transfer (SFPT) \cite{liaoSkeletonfreePoseTransfer2022}. Finally, in the interaction module, the weight editor serves as the central tool for users to tailor the results by adjusting the implicit correspondence between the 3D human character and the target character. Additionally, the pose editor offers flexibility for making minor modifications to human motion.
Based on the proposed framework, we develop an interactive system for stylized 3D character animation.
We perform evaluation studies for our approach. First, a questionnaire study shows evidence that the output after user guidance is more natural and similar to the input video than the automatically generated result. Second, a case study shows the efficiency of our approach in animating stylized 3D characters that meet users' expectations.

In summary, our work makes contributions in the following three aspects. 
\begin{itemize}
    \item An innovative mixed-initiative framework that enables stylized 3D characters to mimic the motion of human videos. 
    \item 
      A prototype system enabling users to guide the creation of stylized 3D character animations. 
    \item 
      A questionnaire study to validate our approach's ability to generate smooth and faithful stylized 3D character animations, and case studies showcasing the framework's utility in creating diverse stylized 3D character animations.
\end{itemize}

\section{Background \& Related Work}\label{sec-2}
In this section, we provide background information on 3D characters and review existing research on animated 3D characters, video-driven character animation, and motion retargeting for 3D characters.

\subsection{Preliminaries on 3D characters} 
The domain of 3D character modeling is a multifaceted discipline that hinges upon the understanding and proficient utilization of foundational constructs. 
Among these constructs, the concept of mesh assumes paramount importance, constituting the cornerstone upon which the intricate artistry of digital character representation is built. 
Mesh can be defined as a structural lattice comprising interconnected vertices, edges, and polygons (faces). It serves as the skeletal framework upon which the aesthetic and functional attributes of a 3D object, such as a character, are meticulously superimposed. 
In essence, mesh encapsulates the character's foundational form, determining its geometric shape, volumetric structure, and spatial orientation. 
Beyond this rudimentary delineation of shape, mesh begets a host of creative opportunities, facilitating the application of intricate details, textures, and animation sequences that breathe vitality into virtual entities. 
In this work, we purely deform stylized 3d characters based on mesh to generate stylized 3d character animation. 
Compared with the traditional method of deforming 3d characters based on skeleton, this method has higher freedom and can meet the requirement of driving characters with various proportions.


\subsection{Animated 3D Characters}

Animated 3D characters add life and personality to digital media \cite{bessmeltsevGesture3DPosing3D2016}
, games \cite{amani3DModelingAnimating2019}
, films \cite{maoSketchingoutVirtualHumans2006}
, and other interactive experiences by portraying characters that can move, emote, and interact in believable ways. Compared with static 3D characters, animated 3D characters are more competent in bringing a higher level of realism, engagement, and storytelling \cite{song3DCharacterAnimation2015}
. Accordingly, there has been a series of works seeking to reduce the difficulty of producing animated 3D characters. There are some representative tools for animating 3D characters. Maya \cite{Maya2019}, one of the most popular and full-featured 3D animation software packages, allows modeling, rigging, animation, simulations, rendering, and more. The open-source 3D creation suite Blender \cite{Blender3DModelling2018} that provides similar functionalities has a strong community and many add-ons.

A number of studies have focused on developing tools for regular users, rather than professional designers, with the primary goal being to simplify the process of setting up animations. 
These works commonly use a probabilistic model to capture limited local variations \cite{liMotionTextureTwolevel2002}
or rely on large motion datasets \cite{holdenDeepLearningFramework2016}
. Among these studies, motion modeling and synthesis have been the most extensively researched areas \cite{holdenDeepLearningFramework2016, liExamplebasedMotionSynthesis2023, jainOptimizationbasedInteractiveMotion2009, habibieRecurrentVariationalAutoencoder2017, lamouretMotionSynthesisExample1996, ravaniMotionSynthesisUsing1983, liGanimatorNeuralMotion2022}
. 
Holden et al. \cite{holdenDeepLearningFramework2016} present a deep learning framework for synthesizing and editing natural character motions using a convolutional autoencoder to learn a motion manifold and a feedforward neural network to map high-level parameters to the manifold.
Habibie et al. \cite{habibieRecurrentVariationalAutoencoder2017}
propose a recurrent variational autoencoder model for human motion synthesis that uses a control signal to generate more accurate long-term predictions. 
However, previous work has primarily focused on generating human character animation using data derived from humans\cite{ofliBerkeleyMhadComprehensive2013, mullerMoCapDatabaseHdm052007}, studies on stylized (both humanoid and non-humanoid) 3D characters remain limited. There is a high demand for stylized 3D characters in today's society, with applications spanning areas such as storytelling\cite{maBuildingCharacterAnimation2003}, games\cite{sloanVirtualCharacterDesign2015}, and education\cite{xiaoAnimationTrendsEducation2013}. Therefore, our work further leverages the abundance of human videos and uses them to animate stylized 3D characters by mimicking the human motion in the videos.


\subsection{Video-Driven Character Animation}
Video-driven character animation is the task of animating characters by the motion in videos \cite{deaguiarVideoDrivenAnimationHuman2007}. 
This task can be exceedingly time-consuming when undertaken manually. Our primary focus is on automating video-driven character animation. Currently, this field can be broadly categorized into two primary types of video-driven character animation: 2D character animation and 3D character animation.

Numerous approaches to video-driven 2D character animation are not domain-specific, and as such, they do not rely on a strong prior knowledge of the animated object. 
Traditional methods \cite{davisSketchingInterfaceArticulated, breglerTurningMastersMotion2002, sykoraAsrigidaspossibleImageRegistration2009} solve this problem by incorporating geometric transformations like affine transformation \cite{breglerTurningMastersMotion2002}. 
Recently, by integrating deep neural networks (DNNs), techniques \cite{siarohinAnimatingArbitraryObjects2019, siarohinFirstOrderMotion2019, suLiveSketchVideodriven2018} for video-driven 2D character animation have emerged. 
For instance, Siarohin et al. \cite{siarohinAnimatingArbitraryObjects2019}
propose Monkey-Net, a deep architecture that decouples appearance and motion information to generate a video in which the target object is animated according to the driving video. Siarohin et al. \cite{siarohinFirstOrderMotion2019}
further introduce FOMM, using a set of self-learned key points together with local affine transformations to tackle the issue of poor generation quality in the case of large object pose changes. 
In comparison, due to the complexity of 3D characters, the task of video-driven 3D character animation is way more intricate.
Video-driven 3D character animation tasks are always based on motion capture models \cite{deaguiarVideoDrivenAnimationHuman2007, dongMotionCaptureInternet2020, baakDatadrivenApproachRealtime2013, kicirogluActiveMoCapOptimizedViewpoint2020, choiStaticFeaturesTemporally2021, wandtCanonposeSelfsupervisedMonocular2021, liLearningHumanDynamics2022}
, i.e., models that capture the motion in videos. 
While the existing body of literature has made significant strides in the field of video-driven 3D character animation, it is clear that there is a crucial gap when it comes to animating stylized 3D characters using videos of characters with diverse proportions, a feat that is attainable for 2D characters. To address this disparity, we present an approach that enables the animation of characters with proportions distinct from those of human characters based on human videos. 

\subsection{Motion Retargeting for 3D Characters}
Motion retargeting is centered around the transfer of movements from a source character to a target character with varying proportions. Within motion retargeting, there exists a sub-task known as pose retargeting, which is the primary focus of this section. In essence, the objective of pose retargeting is to generate a pose for a character with a different body shape, replicating the original pose while maintaining a sense of realism and naturalness.
\cite{mourotSurveyDeepLearning2022} 
. Studies on motion retargeting could be divided into skeleton-based and skeleton-free methods.

For skeleton-based methods, Gleicher \cite{gleicherRetargettingMotionNew1998a} 
is the first one to present a practical technique to retarget motion between characters by posing it as a constrained optimization problem that preserves key features of the original motion while limiting undesirable changes. 
After it, many non-DNNs approaches represented by methods based on physics and kinematics constraints \cite{albornoRobustPhysicsbasedMotion2018, choiOnlineMotionRetargetting, takPhysicallyBasedMotionRetargeting, monzaniUsingIntermediateSkeleton2000, abdul-massihMotionStyleRetargeting2017, bassetContactPreservingShape2020a}
are proposed. 
Recently, as DNNs have advanced, new techniques for motion retargeting built on DNNs have emerged \cite{kimMotionRetargettingBased2020, abermanSkeletonawareNetworksDeep2020, villegasNeuralKinematicNetworks2018, villegasContactAwareRetargetingSkinned2021, limPMnetLearningDisentangled, maheshwariTransfer4DFrameworkFrugal, liLearningSkeletalArticulations2021}
. Most of these methods are recurrent neural networks (RNNs), convolutional neural networks (CNNs), or graph neural networks (GNNs) -based. However, these methods require skeletons of the source and the target characters to be the same or topologically equivalent \cite{abermanSkeletonawareNetworksDeep2020} 
. 
Within skeleton-free methods, predominantly rooted in DNNs, several approaches have been proposed that rely on explicit correspondence between the source and target characters.
However, explicit correspondence limits the ability to transfer pose between distinct characters. Skeleton-Free Pose Retargeting (SFPT) \cite{liaoSkeletonfreePoseTransfer2022} is one of the state-of-the-art and representative works for transferring poses from human characters to stylized 3D characters without skeletal rigging. It utilizes skinning weights for correspondence between the source and the target character rather than explicit correspondence like key points, which provides good flexibility. HMC \cite{wangHMCHierarchicalMesh2023} extends SFPT to motion retargeting by learning the correspondence in a coarse-to-fine fashion by integrating the retargeting process with a mesh-coarsening pipeline. Wang et al. \cite{wangZeroshotPoseTransfer2023} 
present a zero-shot approach that requires only the widely available deformed source non-stylized avatars in training, and deforms target stylized characters of significantly different shapes at inference. In our work, we utilize the strong prior induced by the pre-trained SFPT model for the task of motion transfer for 3D characters.
\section{Design Considerations}\label{sec-3}

Inspired by the desire to streamline the creation of stylized 3D character animations, we explore MoCap and MoTrans techniques.
Instead of adhering to the conventional practices of professional animators, which involve intricate rigging and meticulous skeleton adjustments to animate 3D characters from the ground up, our approach allows users to specify their desired motions based on readily available single-view human videos. 
By leveraging a multitude of human videos, users can harness a broader spectrum of motion possibilities compared to conventional animation methods. 

One major design consideration is to support both direct generation and allow users to adjust according to their expectations (\textbf{C1}). 
Given that MoCap and MoTrans inherently introduce inaccuracies and may not always align with users' expectations, it becomes essential to provide options for modifying the outcomes to better match what users anticipate.
We also endeavor to simplify the intricate and complex elements of deep learning models when users are making modifications (\textbf{C2}). 
Due to the limited interpretability of the adjustable components within these models, we aim to hide the intricate components, enabling users to make modifications without needing direct access to the models' internals.
\section{Framework}\label{sec-4}

In this section, we present a framework for animating a stylized 3D character by mimicking the human motion of a single-view human video.

\subsection{Preliminaries} 
In this subsection, we present the concept of skeleton and skinning weight of 3D characters as preliminaries.

\subsubsection{Skeleton of 3D Characters}
The concept of skeleton in 3D character is fundamental and critical, acting as the invisible framework upon which the character's movements are built.
Just as human-beings' skeletons provide structure and support for bodies, a 3D character's skeleton defines its underlying framework, allowing it to be articulated, animated, and brought to life in a virtual world. 

\subsubsection{Skinning Weight of 3D Characters}
Deformation Primitives are the basic elements used to control and implement shape changes in a 3D character or object, and the process of binding a set of deformation primitives to 3D mesh vertices is called skinning. This allows the movement or deformation of the primitives to correspondingly affect the movement or deformation of the associated vertices, simulating changes in the object, without having to directly manipulate the complex 3D mesh structure. The extent to which each primitive affects the deformation of the character mesh can be controlled by assigning each vertex a corresponding skinning weight. For a mesh with $K$ deformation primitives, the skinning weight of a vertex is a $K$-dimensional vector $W$ satisfying the following constrants:
\begin{equation}
\begin{aligned}
    & ||W||_1 = 1, \\
    & 0 \leq W_i \leq 1, i = 1, 2, ..., K,
\end{aligned}
\end{equation}
where $W_i$ represents the impact of deformation primitive $i$ on the vertex. 
The most commonly used deformation primitive is skeleton. However, skeleton-based deformation requires skeletal rigging. 
Aiming at automatic generation of stylized 3D character animations, we deform the user-input stylized 3D characters based on implicit deformation primitives \cite{liaoSkeletonfreePoseTransfer2022} which is more flexible. 

\subsection{Overview}

\begin{figure*}[h]
  \centering
  \includegraphics[width=\linewidth]{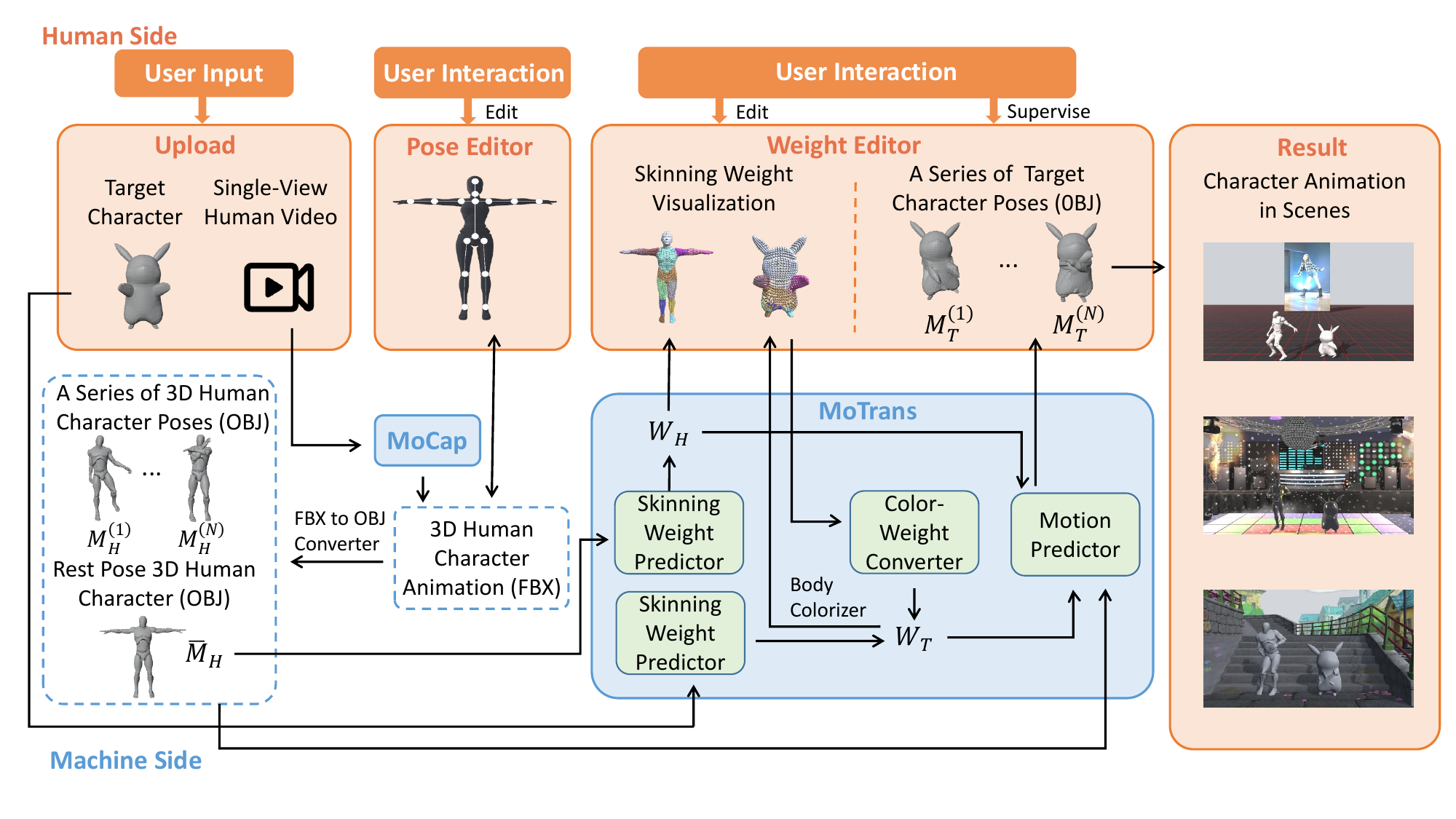}
  \caption{
  Overview of our approach. 
  On the human side, the input consists of a stylized 3D character (the target character) and a single-view human video. The end result is the stylized 3D character animation, which is showcased in specific scenes. Throughout the workflow, users have the option to customize the outcome using both the pose editor and the weight editor.
  On the machine side, the process involves the MoCap module, which captures human motion from the input video. This captured motion can be adjusted using the pose editor. The MoTrans module then transfers this human motion onto the target character. Within this module, skinning weights denoted as $W_H$ (for the human character) and $W_T$ (for the target character) are predicted. 
  $W_T$ can be iteratively modified by users using the weight editor, facilitated by the color-weight converter. 
  This modification is guided by supervising the output of the target character's motion as predicted by the motion predictor.
  }
  \label{fig:workflow}
\end{figure*}

Fig. ~\ref{fig:workflow} illustrates our framework. The computational pipeline takes in a single-view human video and a rest-pose stylized 3D character (the target character) $\bar{\mathbf{M}}_T$ as input, and outputs the animated target character that mimics the human motion of the video. For customization, users can edit the skinning weights $W_T$ of the target character to modify the implicit correspondence between the human character and the target character.

Internally, the input video is first fed into the MoCap model of DeepMotion \cite{DeepMotion2023}
to capture the human motion in the form of a series of frames of posed human character meshes, i.e. $\mathbf{M}_H^{(i)}, i = 1, 2, ..., N$. The human motion is editable through the pose editor. The rest-pose human character mesh $\bar{\mathbf{M}}_H$ and target character mesh $\bar{\mathbf{M}}_T$ are fed into the skinning weight predictor of Skeleton-free Pose Transfer (SFPT) \cite{liaoSkeletonfreePoseTransfer2022} to obtain the skinning weights $W_H$ and $W_T$ of vertices of the meshes, respectively. 
The skinning weights indicate the implicit correspondence between the human character and the target character, and users are allowed to modify the implicit correspondence through the weight editor. 
Subsequently, the skinning weights are utilized in the motion predictor, resulting in the animated 3D character mimicking the human motion in the video in the form of a series of frames of meshes, i.e., $\mathbf{M}_T^{(i)}, i = 1, 2, ..., N$.

\subsection{Motion Capture}
We aim to make stylized 3D characters mimic human motion in videos. 
However, there are currently no ready-made tools that can automatically capture the motion from human videos onto 3D character animation. 
Thus, we need a MoCap model to animate a 3D human character by motion of human videos. This serves as the foundation for our subsequent discussion on transferring this motion to the target character.
In order to make our approach accessible, MoCap models that rely on the most readily available resources are ideal. Amongst all kinds of MoCap models, monocular MoCap models, which take single-view videos as input, are the most suitable choice. Single-view videos offer advantages such as ease of capture without the need for calibration, synchronization, or multiple cameras. By utilizing monocular MoCap models for motion capture in our approach, users are able to make stylized 3D characters mimic the motion in any easily accessible single-view human videos, e.g., videos on YouTube. 

The MoCap model of DeepMotion \cite{DeepMotion2023} is applied for 3D human character animation from human videos in our task. 
This highly realistic model is one of the most powerful and widely-used monocular MoCap models and has shown promising results in applying to tasks like character customization and animation from videos, user-generate games \cite{AvaturnDeepMotion2023, RobloxDeepMotion2023}
. 
Through the DeepMotion API, this model can be leveraged by users by sending a single-view human video and receiving generated 3D human character animation with a simple API call. 
This allows us to tap into the models’ strengths without needing to train or host a Mocap model itself. 
In our work, we leverage the model through the API to automatically animate 3D human characters based on single-view human videos.

\subsection{Motion Transfer for 3D Characters}

Given the 3D human character animation, we need to transfer the motion from it to the target character, which is a MoTrans task.
MoTrans is the task of transferring motion of the source character to the target character. In our work, we focus on transferring motion of the human character to an arbitrary use-input stylized 3D character, which could be either a humanoid or non-humanoid character. Since the motion of a source character is a sequence of static poses, we treat the task of MoTrans as a sequence of pose transfer tasks.
The main problem for pose transfer is the difference in proportions between the source and target characters. To this end, skeleton-based \cite{albornoRobustPhysicsbasedMotion2018, choiOnlineMotionRetargetting, takPhysicallyBasedMotionRetargeting, monzaniUsingIntermediateSkeleton2000, abdul-massihMotionStyleRetargeting2017, bassetContactPreservingShape2020a} and skeleton-free \cite{liaoSkeletonfreePoseTransfer2022, wangHMCHierarchicalMesh2023, wangZeroshotPoseTransfer2023} methods were proposed. Since skeleton-based methods often require strict similarity between the source and target characters but we aim to be able to transfer poses between the human character and a distinct 3D character, we choose to use skeleton-free methods for pose transfer.

SFPT \cite{liaoSkeletonfreePoseTransfer2022} is a pose transfer model that takes the rest-pose human character $\bar{\mathbf{M}}_H$, the posed human character $\mathbf{M}_H$, and the rest-pose target character $\bar{\mathbf{M}}_T$. For each frame of posed human character meshes $\mathbf{M}_H^{(i)}, i = 1, 2, ..., N$, SFPT is applied to transfer the pose of the human character to the target character. As a learning-based method for automatically transferring poses between 3D characters without skeletal rigging, it allows for handling characters with diverse shapes, topologies, and mesh connectivities. The key idea is to represent the characters in a unified articulation model with a set of deformation body parts, i.e., implicit deformation primitives. Each part can be deformed independently to transfer the pose by a rigid transformation based on the predicted skinning weights. The model mainly consists of three components:

\begin{itemize}
    \item Skinning weight predictor $g_H$: A graph convolution network (GCN) that predicts per-vertex skinning weights to segment the mesh into deformation parts.
    \item Mesh encoder $g_e$: A GCN that encodes the pose and shape of input mesh into a latent feature embedding.
    \item Transformation decoder $f$: a Multi-Layer Perceptron (MLP) that predicts part-wise rigid transformations to articulate target character by linear blending skinning (LBS) \cite{kavanDirectSkinningMethods}.
\end{itemize}

In our implementation, we utilized the pre-trained SFPT model trained on three datasets, AMASS \cite{mahmoodAMASSArchiveMotion2019}, Mixamo \cite{Mixamo2022} and RigNet \cite{xuRigNetNeuralRigging2020a}, which has shown state of the art performance in pose transfer. Empirically, simply applying the model sequentially by leveraging $\bar{\mathbf{M}}_H$, $\mathbf{M}_H^{(i)}$, and $\bar{\mathbf{M}}_T$ as input for $i = 1, 2, ..., N$ has the following disadvantages:

\begin{itemize}
    \item There would be inconsistencies between the skinning weights predicted by the sequential models, even though the rest-pose human character $\bar{\mathbf{M}}_H$ and target character $\bar{\mathbf{M}}_T$ and the model weights of the skinning predictor remain unchanged.
    \item Predicting the skinning weights $N$ times is redundant and will require more computing resources.
    \item The user interaction module requires one pair of consistent skinning weights for the human and target character for each pose transfer task.
\end{itemize}

Therefore, the skinning weights $W_H$ and $W_T$ are predicted at first by the skinning predictor $g_H$. The skinning weights are then fed into motion predictors, i.e., the rest part of the sequential SFPT models.

\subsection{User Interaction}
We developed two interaction modules: the pose editor and the weight editor. 
The pose editor supports users to motion of the human character captured from the input video.
The weight editor allows intuitive yet effective manipulation of the implicit correspondence between the human character and the target character by editing vertex colors in the user interface. The modification operation is supported by the coloer-weight converter.

\subsubsection{Pose Editor} 
Results of the Mocap model can often contain inaccuracies due to limitations in the capture process. 
The pose editor enables users to directly adjust the character's skeleton for each frame, allowing for precise refinement of the pose as required. 
By editing the joint rotations and translations, the human character's animation can be polished to correct flaws or exaggerations in the original result.
This pose-by-pose editing gives animators precise control over the final motion. It complements the rough human character animation from Mocap with hand-crafted nuance and style.

\subsubsection{Weight Editor}

The weight editor features modifying the implicit correspondence between the human character and the target character by directly changing the color of vertices in the user interface. The underlying logic is manipulating the skinning weights of edited vertices automatically, which is supported by the color-weight converter. 

The basic idea of the color-weight converter is, for a vertex $v$ edited to be of deformation body part $k$, to assign skinning weights according to vertex set of the same deformation body part $k$:
\begin{equation}
    V_k = \{u | argmax(W_u) = k\}.
\end{equation}
In addition, the vertex set $V_k$ have varying degrees of impact on the new skinning weights $W_v^{(k)}$ of $v$, i.e. $W_v^{(k)}$ is calculated by a normalized weighted average of skinning weights of vertices in $V_k$:
\begin{equation}
\begin{aligned}
    & \tilde{W}_v^{(k)} = \sum_{v_i \in V_k} \alpha_i^{(k)} W_i, \\
    & W_v^{(k)} = \frac{\tilde{W}_v^{(k)}}{||\tilde{W}_v^{(k)}||_2}    
\end{aligned}
\end{equation}
where $W_i$ is the skinning weight of $v_i$, and $\alpha_i^{(k)}$ is calculated by the product of two components $\gamma_i$ and $\eta_i^{(k)}$: 
\begin{equation}
    \alpha_i^{(k)} = \gamma_i \eta_i^{(k)}.
\end{equation} 

Considering that it is often desirable for nearby vertices in a mesh to have similar skinning weights, as it could result in desirable outcomes like smooth deformation, continuity, and reduced artifacts, it is important for the skinning weight of the edited vertex $v$ should depend more on nearby vertices in $V_k$ and less on faraway vertices in $V_k$. 
Inspired by the idea of inverse distance weighting, a type of deterministic method for multivariate interpolation with a known scattered set of points, we define $\gamma_i$ as follows:
\begin{equation}
    \gamma_i = \frac{1}{d(x, x_i)^p},
\end{equation}
where $x$ and $x_i$ are the coordinates of $v$ and $v_i$, respectively, $d(x, x_i)$ is the distance between $x$ and $x_i$, $p$ is the power parameter. In our implementation, we use Euclidean distance between the coordinates of $x$ and $x_i$ for $d(x, x_i)$ and set $p = 1$.

Nevertheless, 
in the practical application of our approach, users typically prefer to edit vertices located either at the boundaries or at a significant distance from the deformation body parts to which they need to assign them. 
Therefore, when there are some vertices in $V_k$ that are near the edited vertex $v$, while mistakenly predicted, the newly assigned skinning weight $W_v^{(k)}$ of $v$ tends to be inaccurate due to the effect of the higher weight $\gamma_i$ for the nearby vertices in $V_k$. To this end, we define $\eta_i^{(k)}$ by the density of $v_i$ in $V_k$ estimated by kernel density estimation (KDE) to lower the impact of the mistakenly predicted and nearby vertices in $V_k$: 
\begin{equation}
    \eta_i^{(k)} = \hat{f}_h^{(k)} (x_i)  = \frac{1}{n_k} \sum_{j = 1}^{n_k} K_h(x_i - x_j) = \frac{1}{n_k h} \sum_{j = 1}^{n_k} K(\frac{x_i - x_j}{h}),
\end{equation}
where $n_k = |V_k|$, $K$ is the kernel — a non-negative function, and $h > 0$ is a smoothing parameter called the bandwidth. In our implementation, We use the Scott estimation to estimate $h$ and choose a Gaussian kernel for $K$:
\begin{equation}
    K(u) = \frac{1}{\sqrt{2\pi h^2}} exp(-\frac{u^2}{2h^2})
\end{equation}

Since SFPT allows some deformation body parts to be degenerate, there could be some parts of the target character that are not covered. 
For the special case when users try to assign the vertex $v$ to a non-covered deformation body part $k'$, we assign the $k'$th value of $W_v$ to 0, which is empirically shown effective.
In the special case where users attempt to assign vertex $v$ to a deformation body part $k'$ that is not currently covered, we assign the $k'$th value of $W_v$ to 0, which is empirically shown effective.

\begin{figure}[h]
  \centering
  \includegraphics[width=\linewidth]{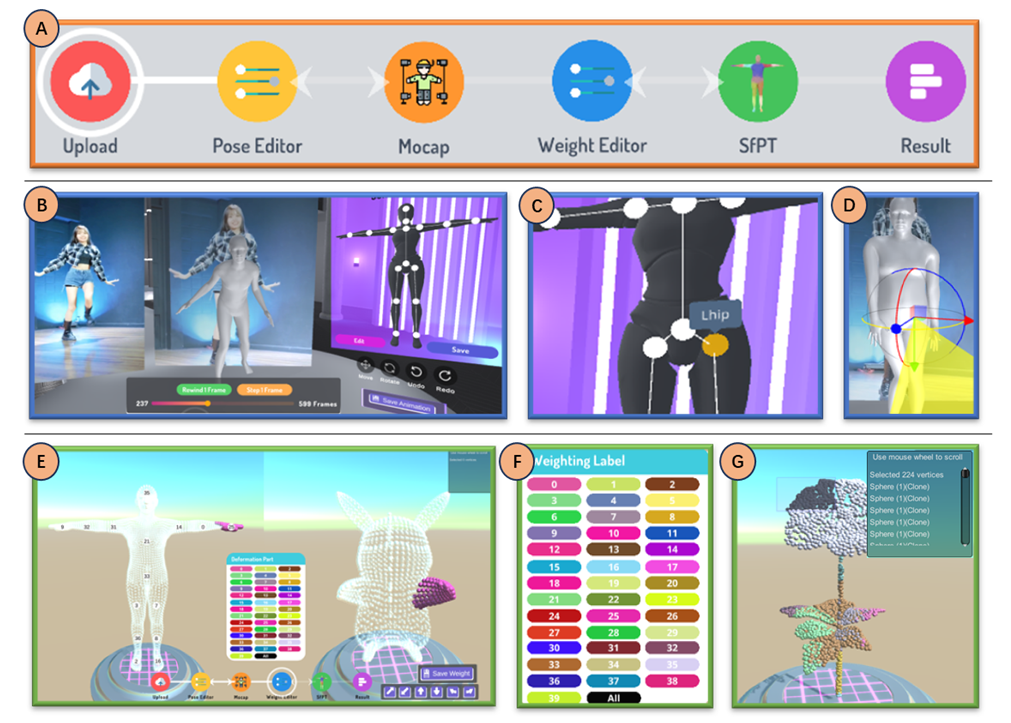}
  \caption{This figure provides an in-depth exploration of VidAnimator's authoring interface, highlighting its essential components: (A) the control bar facilitating seamless scene transitions, (B) the progress bar within the Pose Editor, allowing precise keyframe selection for modification, (C) the interface for selecting character skeletal joints, (D) comprehensive adjustments for selected joints, offering control over six degrees of freedom, (E) a visual representation of selected weight labels in the Weight-Skinning Editor, aiding users in understanding correspondence relationships, (F) the weight label panel, which facilitates the visualization of weight correspondences and vertex color modifications when vertices are selected, and (G) a versatile vertex selection tool for users to create selections, with real-time updates on the count of selected vertices displayed in the upper right panel.}
  \label{fig:system2}
\end{figure}
\section{Authoring Interface}\label{sec-5}

Building upon the proposed framework, we introduce VidAnimator as shown in Fig. ~\ref{fig:teaser}, a Unity-based system that allows users to infuse their personal judgment and creative insights into the generated animations via interactive interfaces. In this section, we will provide a step-by-step introduction to the system's interface and its interactive workflow. 

VidAnimator is primarily composed of six distinct interfaces, thoughtfully structured into four pivotal stages: Upload of Video and 3D Character Data, Video-to-Human Motion Capture, Character-to-Stylized Motion Transfer, and Application of Results in Our Provided Scenes. The seamless transition between these interfaces is facilitated by the control bar, located at the bottom of each interface, as illustrated in Fig.~\ref{fig:system2}(A). This intuitive interface-switching mechanism ensures users are well-informed about their current stage within the animation generation process, enabling them to tailor their experience towards personalized animation creation or interactive refinement. In the subsequent sections, we will provide a detailed introduction to each of these four pivotal stages.

\subsection{Upload of Video and 3D Character Data}

In this initial phase, as depicted in Fig.~\ref{fig:teaser}(A), users begin by uploading video content, which should not exceed 20 seconds in duration, through the "Upload Video" button. After a successful upload, users can then preview the uploaded video. Additionally, users have the option to upload 3D characters in OBJ format using the "Upload Character" button, with the ability to preview these characters upon successful upload. To accommodate users who may not possess their own video or 3D character data, we provide a selection of pre-supplied videos and 3D characters within the interface. These pre-supplied options are readily available for users and are recorded for later use in subsequent stages. Once users select the "MoCap" button from the control bar, the system initiates the motion capture process from the uploaded videos, displaying interim results within the designated views.

\subsection{Video-to-Human Motion Capture}

Within the MoCap interface, as depicted in Fig.~\ref{fig:teaser}(C), users can observe the motion capture (MoCap) results, which prominently display the generated human animation at the center, flanked by the input videos on the left side. Users can seamlessly acquire the human animation generated from their input videos without the need for additional correction. However, for those who seek further refinement or correction of the generated human animation, the "Pose Editor" button within the control bar provides access to the editor mode. Within the pose editor interface as shown in Fig.~\ref{fig:teaser}(B), users can utilize a progress bar (Fig.~\ref{fig:system2}(B)) to navigate through individual video frames, allowing for an in-depth inspection of each frame's content and action. The left side of the progress bar presents the current frame number, while the right side displays the total frame count of the video. Above the progress bar, users will find buttons for navigating forwards or backwards by a single frame. After reviewing the content, users can select keyframes for editing. The right-side panel offers editable joints(Fig.~\ref{fig:system2}(C)), empowering users to fine-tune joint positions and rotation angles, spanning a comprehensive range of six degrees of freedom(Fig.~\ref{fig:system2}(D)). Upon making adjustments for each frame, users can click the "Save" button within the joint selection panel to preserve their modifications. After completing all necessary frame adjustments, users can finalize and save their results by clicking "Save Animation."

\subsection{Character-to-Stylized Motion Transfer}

In this stage, users activate the "MoTrans" button to assess the animation outcomes generated by the generalized model through motion transfer as shown in Fig.~\ref{fig:teaser}(E). The result of the target animation is shown on the right side of the screen, while the human animation from the previous stage occupies the central position, and the video remains on the left side of the screen. To explore the character's quality from various angles, users can employ the two rotation buttons positioned at the bottom right, enabling a versatile view of the character. Additionally, three buttons in the middle provide control over the intensity of lighting for enhanced visualization. If users wish to make additional refinements to the stylized model's motion, they can access the interface by selecting the "Weight Editor" button from the control bar.

Within the Skinning-Weight Editor interface(Fig.~\ref{fig:teaser}(D)), the left side provides a visual representation of all vertex weights of the character model, with distinct colors denoting different weight labels. To enhance user understanding of these color associations, numerical labels corresponding to various parts of the human character have been assigned. The right side displays the visualized results of the vertex weights of the initially uploaded or selected target model. Located in the lower right corner is a control bar featuring six buttons, offering users the capability to explore the model from various angles and positions. The first set of buttons on the left controls forward and backward movement, akin to the "W" and "S" keys on the keyboard. In the middle, the second set manages upward and downward motion, corresponding to the "Q" and "E" keys. Lastly, the rightmost buttons enable left and right rotation of the model, mirroring the functionality of the "A" and "D" keys. The central panel accommodates color-coded buttons(Fig.~\ref{fig:system2}(F)), each associated with a specific label number. These buttons serve a dual function: when no vertices are selected, clicking a color button highlights the corresponding correspondence between vertices in the source and the target character(Fig.~\ref{fig:system2}(E)). However, when vertices are selected, clicking a color button enables users to modify both the color and weight label for the selected vertices.

User interaction primarily centers around the target character on the right side. Holding down the right mouse button and dragging enables users to select multiple vertices by creating a selection rectangle(Fig.~\ref{fig:system2}(G)). Holding the SHIFT key while performing the same action preserves the previously selected vertices while adding the newly selected ones. Single-clicking with the left mouse button allows users to add or remove vertices one by one. The total number of currently selected vertices is displayed in real-time in the upper right corner of the screen for user reference. Selected vertices appear in black until their weight labels are adjusted through the central panel. Upon completing all necessary modifications, users can save the weights by clicking the "Save" button located at the lower right. Subsequently, users can click "MoTrans" to view the results after the modification process.

\subsection{Application of Results in Our Provided Scenes}


Considering the diverse backgrounds of our user base, some users may not be familiar with using OBJ format animation sequences. To face this issue, it's important to provide additional guidance on their effective utilization.
Towards the conclusion of our system, we provide several predefined scenes, allowing users to seamlessly integrate their generated animation results into these environments(Fig.~\ref{fig:teaser}(F)). Users can further enhance their creations through camera movements along predefined paths, enabling video capture to generate videos more easily and facilitating the effortless sharing of their results.

\subsection{Implementation} 
Our system was implemented as a client/server web application. The front end was built with Unity for user interactions. The computational framework for generating animated 3D characters was implemented in Python using Pytorch. The Flask framework is used to handle the messaging between the front end and the back end.
\section{Method Analysis}\label{sec-6}
Our system was implemented as a client/server web application. The front end was built with Unity for user interactions. The computational framework for generating animated 3D characters was implemented in Python using Pytorch. The Flask framework is used to handle the messaging between the front end and the back end.

\section{Method Analysis} 
\label{sec:method_analysis}
\begin{figure*}[h]
  \centering
  \includegraphics[width=\linewidth]{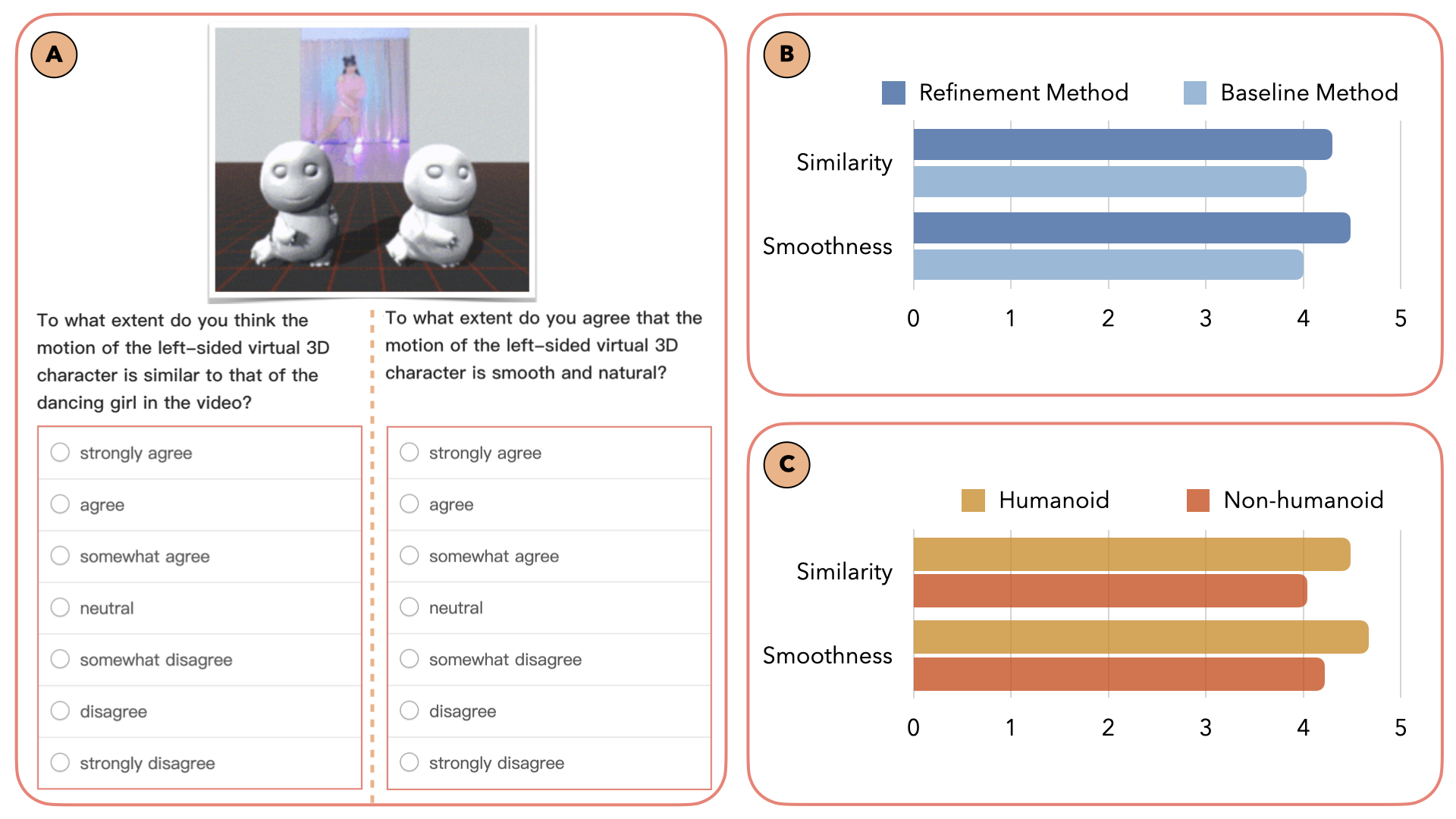}
  \caption{Sample questions and results in Questionnaire Study (N=30): (A) Sample questions specifically designed to evaluate the similarity and smoothness of the animated virtual 3D characters. (B) The average scores from the questionnaire study indicate that our refined method outperforms the baseline approach in both similarity and smoothness. (C) A breakdown of the average scores for each category of characters reveals that humanoid 3D characters surpass non-humanoid 3D characters in performance.}
  \label{fig:Qstudy}
\end{figure*}

\subsection{Questionnaire Study}
We conducted a questionnaire-based study to assess the effectiveness of our proposed framework. More specifically, through the study, our objectives were to determine: 1) whether our complete framework demonstrably improves the quality of motion mimicry of virtual 3D characters, and 2) whether our proposed framework is universally applicable to different categories of stylized virtual 3D characters.

\subsubsection{Setup}
The questionnaires were distributed through the ``WenJuanXing" platform. We organized two virtual 3D character groups: the humanoid group (e.g., Charizard), and the non-humanoid group (e.g., Flower). The humanoid group comprises three distinct cases, while the non-humanoid group comprises four distinct cases. To ensure the objectivity of the questionnaire results, the order of questions within each case were randomized. We converted the initial input human videos and output animated 3D characters into GIF formats to enhance the synchronization of the videos within each study case.

\subsubsection{Study Process}
The questionnaire study is designed to evaluate the effectiveness of our proposed framework in enabling virtual 3D characters to mimic the movements of human characters. The primary objective of our pipeline is to demonstrate the similarity in movements between the input and output characters, with the smoothness of the movements in the output character serving as a secondary criterion. To this end, we first ask users to rate the similarity of movements between the initial input of the human character and the output virtual 3D character. Subsequently, users are asked to rate the smoothness and naturalness of the movements exhibited by the output virtual 3D character.

To enhance user comprehension and readability, the structure of our questions is designed to be clear and straightforward. As sampled in Fig.~\ref{fig:Qstudy}(A), the first question asks, ``To what extent do you think the motion of the left/right-sided virtual 3D character is similar to that of the dancing girl in the video?" The second question inquires, ``To what extent do you agree that the motion of the left/right-sided virtual 3D character is smooth and natural?" We have established two sets of questions for each character: one evaluates the baseline method, and the other focuses on refinement method. The baseline method employs only MoCap and MoTrans, while the refinement method uses a complete framework that includes interactive adjustments. For each question, the input video of the human character is placed in the center, and animated virtual 3D characters created using either the baseline or refinement method are randomly positioned on the left or right sides. Participants are required to rate the similarity and naturalness of these characters based on a 7-point Likert scale, ranging from 0 (strongly disagree) to 6 (strongly agree). Each set of questions contains two individual questions designed to assess similarity and naturalness, respectively. In the final section of our questionnaire study, each participant is asked to respond to an open-ended question. Within the context of this question, participants are invited to share their detailed feedback on virtual 3D characters that have made a significant impression on them, highlighting both the strengths and weaknesses of these animated virtual 3D characters. Overall, our questionnaire study consists of 29 questions.

\subsubsection{Participants}

We recruited 30 participants by disseminating advertisements and questionnaire links on WeChat, a prominent social media platform. The participants are undergraduates, master's students, and Phd students from Chinese universities, specializing in Data Science and Journalism and Media Studies. Upon completing the questionnaire, each participant received a compensation of 6 RMB. Of the 30 participants who took part in the survey, 13 were female and 17 were male, with ages ranging from 18 to 26 years old. Every participant had previously watched 3D animated cartoons and consequently had some level of familiarity and impressions regarding the motion of virtual 3D characters in animated films.

\subsubsection{Effectiveness}
On average, participants took 10 minutes to complete 29 questions (std=6.5, range from 1.8 to 30.5 minutes). We collected 30 questionnaire responses, all of which were deemed valid. Participants generally agreed that our framework effectively allows virtual 3D characters to mimic the motion of human characters in video inputs, rendering these movements in a smooth and natural manner. Notably, when interactive adjustments were applied, the refined virtual 3D characters were perceived to be more similar in movement, as well as smoother and more natural. For our complete framework with interactive refinement, the average score for similarity was 4.30 (std=0.65), and for smoothness and naturalness, the average score was 4.48 (std=0.63). Both scores notably surpassed 4 points, aligning with `somewhat agree' on a 7-point Likert scale. In contrast, using only the baseline methods of MoCap and MoTrans yielded an average similarity score of 4.03 (std=0.73) and an average score for smoothness and naturalness of 4.00 (std=0.82), corresponding to `neutral' on the 7-point Likert scale. As evidenced in Fig.~\ref{fig:Qstudy}(B), compared to the baseline methods, our refinement method achieved significantly higher scores in both similarity, and smoothness and naturalness metrics (significance level $\alpha = 0.05$, paired t-test). This indicates that the interactive adjustments enhance the framework's ability to democratize the creation of stylized 3D character animation by improving the characters' mimicry skills and generating smoother, more natural movements.

In summary, our questionnaire study validated that the inclusion of interactive adjustments in the refinement method yields superior performance in generating stylized 3D character animation compared to using only MoCap and MoTrans. This is evidenced by the simultaneous improvement in the similarity of movements between the virtual 3D characters and their human counterparts in video inputs, as well as in the animated smoothness and naturalness of the virtual characters. These findings corroborate the effectiveness and applicability of our complete framework for generating motion-mimicking stylized virtual 3D characters.
\subsubsection{Comparison of Creating Stylized Animation of 3D Characters with Different Categories}

In our questionnaire study, we examined two distinct categories of virtual 3D characters: the humanoid group and the non-humanoid group. Characters in the humanoid group exhibit body structures closely resembling those of humans, their key joints exhibit a high degree of correspondence with human characters. For instance, the wings of a Charizard can be aligned with human arms, and its legs match up with human legs. Conversely, characters in the non-humanoid group diverge significantly from human character. A prime example would be a flower 3D character, composed solely of stems, leaves, and petals, which cannot correspond to human joints.
Participants generally felt that humanoid virtual 3D characters were more adept at mimicking human movements and displayed a smoother and more natural motion. As shown in Fig.~\ref{fig:Qstudy}(C), the average score for similarity in humanoid 3D characters was 4.48, while the score for smoothness and naturalness stood at an impressive 4.67. In contrast, non-humanoid 3D characters lagged behind, with average scores of 4.04 for similarity and 4.22 for smoothness and naturalness. Participants' feedback highlighted that the humanoid 3D characters were well-received mainly because their articulation points closely resembled human joints, enabling a more accurate mimicry of human movements. On the other hand, non-humanoid 3D character animations were generally described as interesting and novel, 
yet overly abstract. 
Users could discern that these characters attempted to imitate human motion curves, but the absence of directly comparable joints hindered them from convincingly mimicking human movements, also resulting in less smooth and natural motion.

In summary, our study validates the abilities of both humanoid and non-humanoid 3D characters in mimicking human-like movements and achieving smooth and natural motions. Due to the high resemblance to human articulation, humanoid 3D characters demonstrated superior capabilities in mimicry and exhibited more smooth and lifelike motions. 
The abstract nature of non-humanoid 3D characters, however, limited their appeal and performance. 
As a result, our ongoing research aims to further explore the characteristics of non-humanoid 3D characters and experiment with enhancing their capabilities to mimic human-like movements, thereby mitigating the limitations posed by their inherent abstractness.

\subsection{Guidelines} 


Through our continuous experimentation, we have distilled several effective methods for modifying skinning weights. Building upon the findings presented in the previous subsection, we have formulated two guidelines for adjusting skinning weights and provided practical recommendations for users to implement these modifications.

\textbf{G.1: Enhancing Motion Smoothness and Weight Alignment.} In the realm of refinement, our first guideline revolves around enhancing motion smoothness and fluidity. Instead of conducting a comprehensive evaluation of the entire target character, we direct our focus to specific regions where motion replication may require improvements. For humanoid models, renowned for their left-right symmetry, we leverage insights obtained from weight correspondences on one side to enhance the other. In cases where weight labels not found in human anatomy are assigned to stylized characters, deformations may arise, leading to suboptimal results. To address this, we recommend adjusting these weights to align with their counterparts in the human body.

\textbf{G.2 Fostering Abstraction Awareness for Non-Humanoid Characters.} Our second refinement guideline is designed to address the unique challenges faced by non-humanoid characters, as illuminated in our questionnaire study. While these characters are intriguing, their ability to mimic human-like movements is hindered by their abstract nature. We observe that they often lack direct correspondences with human proportions and display asymmetry when compared to generalized models. Consequently, users encounter a higher degree of subjectivity in both the evaluation and refinement processes.

To overcome this challenge, our goal is to enhance users' understanding of the inherent abstraction present in AI-generated correspondences. Users can grasp these correspondences through multiple interactions and trials, gradually developing an intuitive understanding of how AI generates these mappings. This approach empowers users to make informed weight modifications that align with the specific characteristics of each target character input.

In summary, our guidelines provide practical guidance for refining skinning weights, catering to the specific requirements of both humanoid and non-humanoid characters. By following these guidelines, users can further optimize the motion mimicry of virtual 3D characters. The next section will expand on our case study to provide concrete support for our observations.

\begin{figure}[h]
  \centering
  \includegraphics[width=\linewidth]{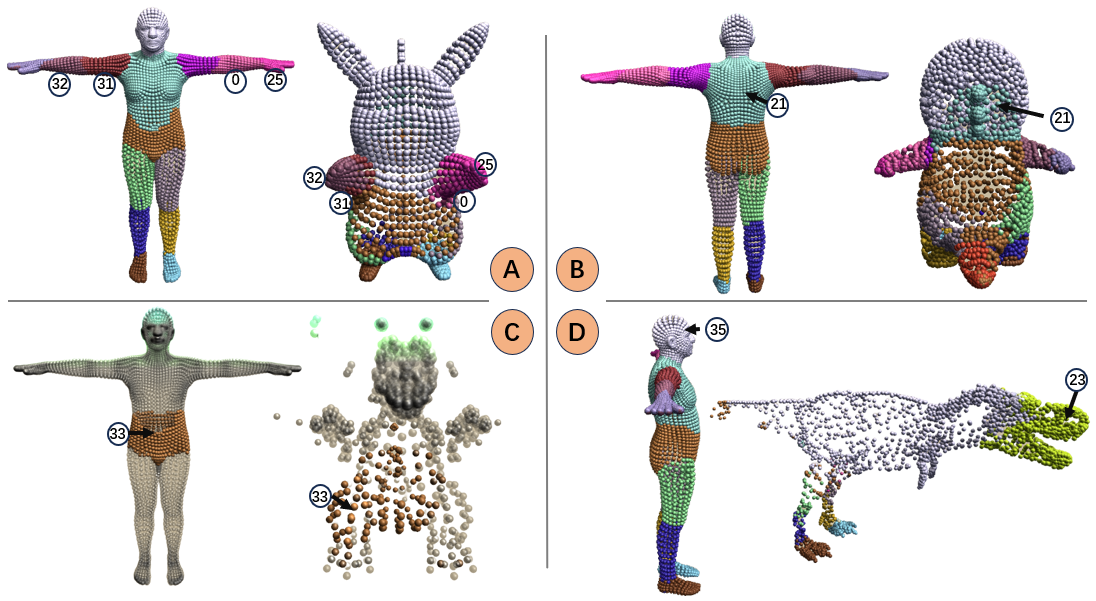}
  \caption{This figure displays the color-coded weight correspondences between the human character and the target characters. Each color corresponds to a specific weight label, highlighting the relationships between different body parts.}
  \label{fig:case_label}
\end{figure}

\subsection{Case Study}

To delve into our experimental findings and address the previously outlined guidelines, we will now embark on a comprehensive case study within the VidAnimator system. In this case study, we will examine three distinct cases: the first two align with G.1, focusing on enhancing motion smoothness, while the third one relates to G.2, with the goal of facilitating users' comprehension of the relationship between human and abstract character correspondences and guiding them in making necessary adjustments.

\subsubsection{Improving Symmetric Motion Smoothness in Humanoid Characters (Pikachu, Nailong, Charizard)}

\begin{figure*}[h]
  \centering
  \includegraphics[width=\linewidth]{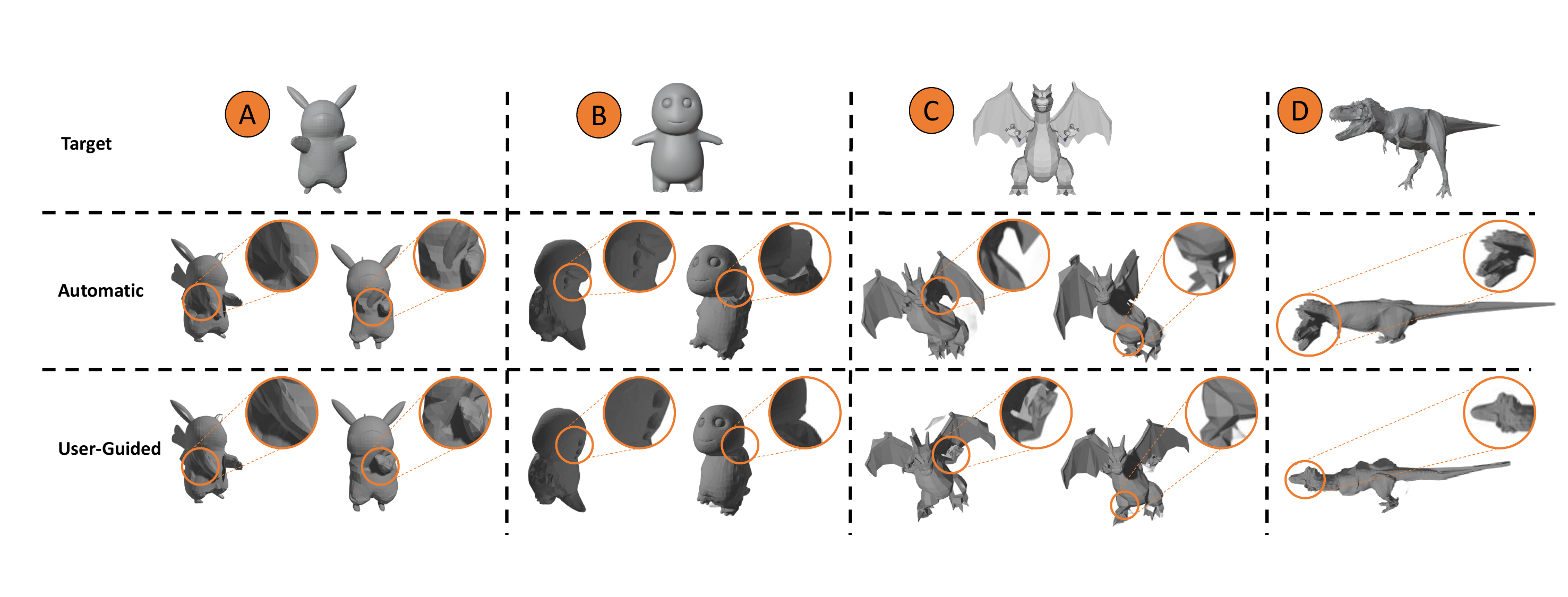}
  \caption{Visualization results of motion transfer: Columns from left to right represent different pairs of characters for motion transfer, namely (A) Pikachu, (B) Nailong, (C) Charizard, and (D) Dinosaur. Rows from top to bottom illustrate the target character's rest pose, automatic animation generation without weight label modification, and user-guided animation with modifications.}
  \label{fig:case12}
\end{figure*}

In this case, we focus on enhancing motion smoothness by facilitating cross-learning between the side of the target character with less accurate motion replication and the better-performing side. Our exploration delved into both left-right and front-back learning relationships. For our investigation, we let users select three different humanoid target characters to showcase our findings, all utilizing the same input video of human dancing, titled 'LastDance.' Our experimental results are presented in Fig. ~\ref{fig:case12}(A)(B)(C).

\textbf{Pikachu} Our first character for animation generation is Pikachu. Initially, noticeable deformations occurred in its right hand, while the movements of the left hand were replicated with higher quality. Investigation into the weight correspondences revealed that the left hand was learning from the human's left forearm (label 0) and left palm (label 25). In contrast, the right hand correlated with the human's right upper arm (label 31) and right forearm (label 32) (see Fig. ~\ref{fig:case_label}(A)). To address this issue, the user optimized the right hand by adjusting it to learn from the right forearm (label 32) and right palm (label 9). This modification significantly mitigated the initial deformation issues (see Fig. ~\ref{fig:case12}(A)).

\textbf{Nailong} Our second humanoid character, Nailong, exhibited deformations in the posterior part of the character's head. To rectify this concern, the user made an adjustment by aligning the weight label with the human head (label 35) rather than the upper body chest (label 21) (see Fig. ~\ref{fig:case_label}(B)). This modification led to a significant improvement in deformation, effectively resolving the issue (see Fig. ~\ref{fig:case12}(B)).

\textbf{Charizard} As our third humanoid character, Charizard displayed animation issues, including excessive deformation and significant left-right oscillations in its right leg. Investigation into the weight correspondences revealed that Charizard's left leg was correlated with the human's left palm (label 25) and left sole (label 16). In contrast, the right leg correlated with the human's lower abdomen (label 33, as seen in Fig. ~\ref{fig:case_label}(C)). To rectify this, the user implemented a modification where the right leg learned from the corresponding weight labels of the left leg, resulting in a noticeable reduction in undesirable oscillations and deformations. Additionally, the user optimized the left hand by aligning it with the human's lower arm instead of only corresponding to the palm, resulting in improved animations, as depicted in Fig. ~\ref{fig:case12}(C).

These findings demonstrate the effectiveness of enhancing motion smoothness and fluidity by leveraging the relationships between different parts of the body in humanoid characters, leading to significant improvements in motion replication quality.

\subsubsection{Adapting Weight Labels Absent in Human Anatomy: A Case Study with a Dinosaur}

In this scenario, our primary concern revolved around weight labels that were absent in the human character's skinning proportions but were present in the target character's label set. Such correspondences frequently resulted in significant deformations within the model. For this particular case study, the user selected a dinosaur model as our target character. During the preview phase, the user observed substantial deformations in the dinosaur's head. An analysis of the weight labels revealed that the dinosaur's head contained label 23, which was not present in the human model (see Fig. ~\ref{fig:case_label}(D)). To address this issue, the user made a critical adjustment by aligning the weight label for the dinosaur's head with that of the human head (label 35). This modification resulted in a noteworthy reduction in deformation, effectively resolving the problem (see Fig. ~\ref{fig:case12}(D)).

\begin{figure}[h]
  \centering
  \includegraphics[width=\linewidth]{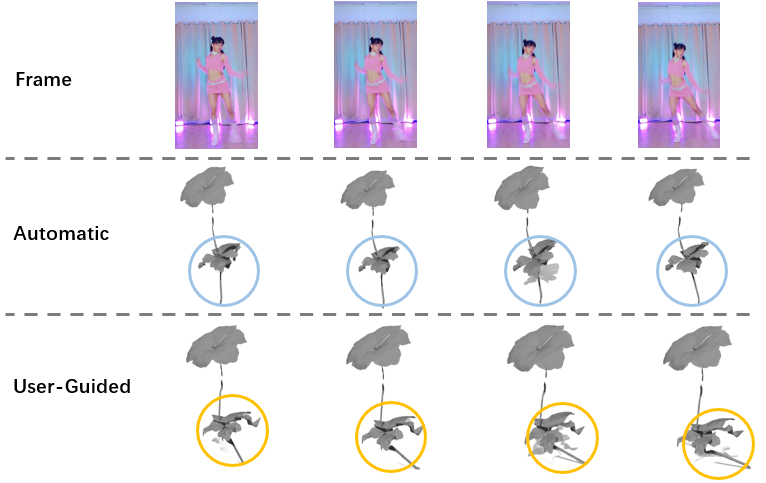}
  \caption{Four consecutive frames illustrating the comparison of movement refinement for the flower model, aligning it with the user's expectations to transform the flower's dance into a joyous and lively performance.}
  \label{fig:case3}
\end{figure}

\subsubsection{Animating Abstract Characters with Unique Morphologies: A Case Study with a Flower}

In this specific case study, the user chose a generalized model that embodies a stylized character, that of a flower, and a select video named "FIFTYFIFTY" as input. This character possesses inherently non-humanoid features, characterized by an asymmetrical morphology. These unique attributes presented a distinctive challenge in establishing coherent correspondences or logical connections with our target character. Upon initially generating animations without any weight adjustments, the user observed a significant disparity: the flower's stem and leaves lacked the vibrant movements seen in the human dance video.

To infuse more vitality into the flower's stem and leaves and enhance their overall liveliness, the user delved into the color-coded weight correspondences. It became evident that the initial correspondences for the flower's stem and root were primarily tied to the human's right leg. However, a detailed analysis of the specific video segment revealed that the human character exhibited more dynamic movements with her left leg. Consequently, the user made a pivotal adjustment, realigning the correspondences with the left leg (label 3, 36, 2). Furthermore, the user identified that the flower's leaves had certain correspondences that exclusively referred to the human's palm and lower body. To improve these correspondences, they were modified to represent the 'upper arm' and 'lower arm' instead. The results, as depicted in Fig. ~\ref{fig:case3}, clearly illustrate the stem mimicking the movement of the human's left leg, while the leaves mirror the human's arm movement. These thoughtful modifications yielded remarkable outcomes, transforming the flower's dance into a joyous, graceful performance that seamlessly harmonized with the user's initial expectations.

User feedback confirms that the VidAnimator system effectively assists users in generating their desired animations, seamlessly aligning with our design considerations (\textbf{C1}). Despite most users lacking in-depth knowledge of the internal mechanisms, our system effectively overcomes challenges introduced by deep learning models during the animation generation process, leading to highly favorable results. This robust performance further reinforces our design considerations (\textbf{C2}). Notably, one user expressed admiration for the selected tools within the Weight Editor, emphasizing their similarity to familiar operations, which significantly enhance the system's usability. However, the user also suggested that the Pose Editor interface could be challenging to navigate and might require excessive manual intervention. Overall, users found the system to be a valuable asset in their animation creation process, highlighting its potential to bridge the gap between user intent and AI-driven animation generation.

\section{Discussion}\label{sec-7}
In this section, we will offer insights into the implications arising from our research, acknowledge the limitations inherent in our study, and explore potential avenues for future investigations.

\subsection{Implication}

\textbf{Advancing Mixed-Initiative Animation Design.} Our work introduces an innovative mixed-initiative framework that enables stylized 3D characters to replicate human motion from videos. The robustness of this pipeline is evident in the statistically significant similarity results from the questionnaire study. By incorporating this efficient pipeline, we advance AI-human collaboration in content creation and unlock new creative possibilities by blending human and AI creativity.

\textbf{Automatic Animation Authorship.} VidAnimator is designed to streamline the animation authoring process, making it more accessible even to users with minimal knowledge of 3D content creation. Our efficient approach, validated through three case studies and the weight label correspondence guidelines, reduces the time and effort required. This reduction in complexity diminishes the entry barrier for individuals interested in creating stylized 3D character animations. As a result, our automatic approach expands the pool of potential creators, fostering a more inclusive and diverse animation community.

\textbf{Enhancing Animation Quality and Smoothness.} Through the questionnaire study and three case studies, we've demonstrated our proficiency in generating smooth and faithful stylized 3D character animations. This has broader implications for animation quality across various domains. Creators can enhance animation quality and smoothness by utilizing our framework and guidelines, ensuring their results align with artistic vision and meet user expectations.

\subsection{Limitations \& Future Work}
\textbf{User-friendliness when the target character is highly abstract. } 
Our approach is capable of Creating stylized 3D character animation from human-videos. As mentioned in Section \ref{sec:method_analysis}, users may encounter challenges when dealing with target characters that diverge significantly from the human character, such as flowers, as establishing a correspondence relationship between them can be non-trivial, even for human-beings.
Future works may explore using a wider variety of video content beyond human videos to animate stylized 3D characters by replacing the MoCap model with other models. 

\textbf{Modification of MoCap results. } Since motion is composed of a series of pose frames, 
the process of substantially altering the motion output of the MoCap model can become inefficient when it necessitates editing many frames consecutively. 
Recent advances in motion synthesis \cite{liExamplebasedMotionSynthesis2023, habibieRecurrentVariationalAutoencoder2017} may help to simplify the process of modifying MoCap results at the cost of sacrificing the high degree of freedom provided by the pose editor. 

\textbf{The Penetration Problem. } 
Penetration problems \cite{liaoSkeletonfreePoseTransfer2022} can arise in the result, especially for chubby target characters. When the shape of the target character is very different from that of the human character of the MoCap model output, the relative positions of the two characters' limbs are different. 
For instance, consider Nailong. When applying a pose where a human character's arms are tightly pressed against the torso, this can result in the arms intersecting with Nailong's body due to his pronounced abdomen and thick waist. 
In future research, it could be worthwhile to investigate the integration of an automated process for selecting a human character for MoCap output from a predefined base of human characters. This selection could align the general physique of the chosen human character more closely with that of the target character. 
\section{Conclusion} \label{sec-8}

In this work, we have introduced an innovative framework and an interactive system for the automated generation of stylized 3D character animations that mimic motions in human videos. Furthermore, our approach enables user modification to customize the results.
Specifically, our approach combines motion capture, motion transfer, and two interaction modules - the pose editor and the weight editor.
The pose editor allows users to refine the human motion captured from videos through skeletal joint adjustments on a frame-by-frame basis. 
The weight editor serves as an intuitive interface that enables users to tailor the motion transfer results. 
This is achieved through indirect modifications to the vertex correspondences between the 3D human character and the stylized 3D character, using skinning weights. Users can accomplish this adjustment by making changes to vertex colors within the interface. 
We evaluated our system through a questionnaire study and case studies. 
The questionnaire study validated that our approach is capable of generating natural and faithful result. 
The case studies demonstrated the utility in creating various stylized 3D character animations. 

\bibliographystyle{ACM-Reference-Format}
\bibliography{main}

\end{document}